\documentstyle[twocolumn,aps,prc,epsf]{revtex}
\draft
\begin{document}

\wideabs{
%
%	title
\title{Centrality Dependence of Kaon Yields in Si+A and Au+Au Collisions 
	at Relativistic Energies}
%
% %%%%%%%%%%%%%%%%%%%%%%%%%%%%%%%%%%%%%%%%%%%%%%%%%%%%%%%%%%%%%%%%%%%%%
%
%	authors
%
\author{(The E--802 Collaboration)}
\author{
                  L.~Ahle,$^{(10,\dag\dag\dag)}$
                 Y.~Akiba,$^{(6)}$
             K.~Ashktorab,$^{(2)}$
                 M.~Baker,$^{(10,\S\S\S)}$
                D.~Beavis,$^{(2)}$
                 P.~Beery,$^{(4)}$
              H.~C.~Britt,$^{(9)}$
                B.~Budick,$^{(11)}$
                 J.~Chang,$^{(4)}$
               C.~Chasman,$^{(2)}$
                  Z.~Chen,$^{(2,\ddag\ddag)}$
                C.~Y.~Chi,$^{(5)}$
                Y.~Y.~Chu,$^{(2)}$
             V.~Cianciolo,$^{(10,**)}$
               B.~A.~Cole,$^{(5)}$
           J.~B.~Costales,$^{(10)}$,
           H.~J.~Crawford,$^{(3)}$
            J.~B.~Cumming,$^{(2)}$
                 R.~Debbe,$^{(2)}$
	     J.~C.~Dunlop,$^{(10)}$
              W.~Eldredge,$^{(4)}$
              J.~Engelage,$^{(3)}$
               S.~Y.~Fung,$^{(4)}$
             J.~Gaardhoje,$^{(12)}$
                 M.~Gonin,$^{(2)}$
                S.~Gushue,$^{(2)}$
              H.~Hamagaki,$^{(15)}$
                A.~Hansen,$^{(12)}$
                L.~Hansen,$^{(12)}$
                O.~Hansen,$^{(12)}$
             R.~S.~Hayano,$^{(13)}$
               S.~Hayashi,$^{(2)}$
           G.~Heintzelman,$^{(10)}$
                 S.~Homma,$^{(6)}$
                  E.~Judd,$^{(10,\ddag\ddag\ddag)}$
                H.~Kaneko,$^{(8)}$
                  J.~Kang,$^{(4,***)}$
               S.~Kaufman,$^{(1)}$
              W.~L.~Kehoe,$^{(10)}$
               A.~Kumagai,$^{(14)}$
                K.~Kurita,$^{(5,\S\S)}$
             R.~J.~Ledoux,$^{(10)}$
             M.~J.~LeVine,$^{(2)}$         
                  J.~Luke,$^{(9)}$
                 Y.~Miake,$^{(14)}$
           D.~P.~Morrison,$^{(10,\S\S\S)}$
              R.~J.~Morse,$^{(10)}$
             B.~Moskowitz,$^{(2)}$
               M.~Moulson,$^{(5)}$
              S.~Nagamiya,$^{(5,\S)}$
         M.~N.~Namboodiri,$^{(9)}$
              T.~K.~Nayak,$^{(5)}$
               C.~A.~Ogilvie,$^{(10)}$
                J.~Olness,$^{(2)}$
            C.~G.~Parsons,$^{(10)}$
           L.~P.~Remsberg,$^{(2)}$
              D.~Roehrich,$^{(2,\ddag)}$
            P.~Rothschild,$^{(10)}$
                  H.~Sako,$^{(15)}$
               H.~Sakurai,$^{(13)}$
           T.~C.~Sangster,$^{(9)}$
                  R.~Seto,$^{(4)}$
               K.~Shigaki,$^{(6,\dag\dag)}$
                 R.~Soltz,$^{(10,\dag\dag\dag)}$
               P.~Stankus,$^{(5,**)}$
           S.~G.~Steadman,$^{(10)}$
        G.~S.~F.~Stephans,$^{(10)}$
               T.~W.~Sung,$^{(10)}$
                Y.~Tanaka,$^{(7)}$
         M.~J.~Tannenbaum,$^{(2)}$
             J.~H.~Thomas,$^{(9,*)}$
              S.~R.~Tonse,$^{(9,*)}$
          S.~Ueno-Hayashi,$^{(14)}$
           J.~H.~van Dijk,$^{(2)}$
           F.~Videb$\ae$k,$^{(2)}$
              O.~Vossnack,$^{(5,\dag)}$
            V.~Vutsadakis,$^{10)}$
                  F.~Wang,$^{(5,*)}$
                  Y.~Wang,$^{(5)}$
             H.~E.~Wegner,$^{(2)}$
              D.~Woodruff,$^{(10)}$
                    Y.~Wu,$^{(5)}$
                 G.~H.~Xu,$^{(4)}$
                  K.~Yagi,$^{(14)}$
                  X.~Yang,$^{(5)}$
               D.~Zachary,$^{(10)}$
               W.~A.~Zajc,$^{(5)}$
                   F.~Zhu,$^{(2)}$
and                 Q.~Zhu$^{(4)}$
}

%
%	institutions
%
\bigskip

\address{
$^{(1)}$ Physics Division, Argonne National Laboratory, Argonne,
        Illinois 60439-4843\\
$^{(2)}$ Brookhaven National Laboratory, Upton, New York 11973\\
$^{(3)}$ Space Sciences Laboratory, University of California,
        Berkeley, California 94720\\
$^{(4)}$ University of California, Riverside, California 92507\\
$^{(5)}$ Columbia University, New York, New York 10027\\
        and Nevis Laboratories, Irvington, New York 10533\\
$^{(6)}$ High Energy Accelerator Research Organization (KEK), Tanashi-branch,
(Tanashi) Tokyo 188, Japan\\
$^{(7)}$ Kyushu University, Fukuoka 812, Japan\\
$^{(8)}$ Kyoto University, Sakyo-Ku, Kyoto 606, Japan\\
$^{(9)}$ Lawrence Livermore National Laboratory, Livermore, California
        94550\\
$^{(10)}$ Laboratory for Nuclear Science, Massachusetts Institute
        of Technology, Cambridge, Massachusetts 02139\\
$^{(11)}$ New York University, New York, New York 10003\\
$^{(12)}$ Niels Bohr Institute for Astronomy, DK-2100 Copenhagen
        0, Denmark\\
$^{(13)}$ Department of Physics, University of Tokyo, Tokyo 113,
        Japan\\
$^{(14)}$ University of Tsukuba, Tsukuba, Ibaraki 305, Japan\\
$^{(15)}$ Center for Nuclear Study, School of Science, University of Tokyo, 
Tanashi, Tokyo 188, Japan\\
}

%\date{\today}
\maketitle

%%%%%%%%%%%%%%%%%%%%%%%%%%%%%%%%%%%%%%%%%%%%%%%%%%%%%%%%%%%%%%%%%
%
%	abstract
%
\bigskip
\begin{abstract}
Charged kaon production has been measured in Si+Al and Si+Au
collisions at 14.6$A$ GeV/$c$, and Au+Au collisions at
11.1$A$ GeV/$c$ by Experiments 859 and 866 (the E--802 Collaboration)
at the BNL AGS.
Invariant transverse mass spectra and rapidity distributions for
both $K^+$ and $K^-$ are presented.
The centrality dependence of rapidity-integrated kaon yields is studied.
Strangeness enhancement is observed as an increase in 
the slope of the kaon yield with the total number of participants as well as 
the yield per participant.
The enhancement starts with peripheral
Si+Al and Si+Au collisions (relative to N+N) and appears to saturate
for a moderate number of participating nucleons in Si+Au collisions.
It is also observed to increase slowly with
centrality in Au+Au collisions, to a level in the most central Au+Au
collisions that is greater than that found in central Si+A collisions.
The enhancement factor for $K^+$ production are
 $3.0\pm 0.2 {\rm (stat.)} \pm 0.4 {\rm (syst.)}$ and
 $4.0\pm 0.3 {\rm (stat.)} \pm 0.5 {\rm (syst.)}$,
respectively, for the most central 7\% Si+Au collisions and
the most central 4\% Au+Au collisions relative to N+N at the
correponding beam energy.
\end{abstract}


}	%wideabs

%%%%%%%%%%%%%%%%%%%%%%%%%%%%%%%%%%%%%%%%%%%%%%%%%%%%%%%
%
%	introduction
%
\section{Introduction}

Strange particle production has long been considered as a useful
experimental probe of heavy-ion collisions because
strange particle yields are thought to be sensitive to physics resulting from the high baryon densities
attained in heavy-ion collisions at AGS energies~\cite{Ahl98:proton}.
In particular, quark-gluon plasma formation is expected to
result in enhanced strangeness production
\cite{Koc83:Strange,Koc86:Probing,Raf82:Formation,Raf82:Strangeness,Hei87}.
Such an enhancement was observed by BNL Experiment 802 in
Si-induced collisions as an increase in the $K^+$ yield in
central collisions relative to the pion
yield~\cite{Abb90:Kaon,Abb92:centrality}.
However, secondary interactions among hadrons
have been proposed to explain the observed strangeness enhancement,
and cascade models implementing these secondary interactions
can approximately reproduce the kaon yields measured in
Si+A collisions~\cite{Mat89:K,Sor91:rqmd,Kah92:strangeness,Li95:dense}.
Because there are, in principle, many
mechanisms that can increase the yield of strange particles in
heavy-ion collisions, the process(es) responsible for the observed
strangeness enhancement have not experimentally
been clearly identified. However,
with the present high statistics Si+A and Au+Au data,
the systematics of strangeness production can now be studied
as a function of the system size from very small systems to 
very large systems, thus, 
providing more stringent constraints on theoretical
explanations of strangeness enhancement.

  In this paper, semi-inclusive $K^+$ and $K^-$ spectra
measured by E--859 in Si+Al and Si+Au collisions at
14.6$A$ GeV/$c$ and by E--866 in Au+Au collisions at
11.1$A$ GeV/$c$ at the BNL AGS are presented.
Both the Si+A and Au+Au measurements were made by the E--802
collaboration using the Henry Higgins spectrometer~\cite{Abb90:Single}.
Previous measurements of kaon
production in Si+A collisions have been presented by
E--802~\cite{Abb90:Kaon,Abb92:centrality,Abb94:Charged}.
Kaon data from Au+Au collisions at slightly higher beam energy are 
presented by E--866 in~\cite{Ahl98:kaon_auau}.

This paper is organized in the following way.
First, descriptions of the experimental apparatus and data collection
are given in section~II. 
Then, the data analysis is described in section~III.
Results on kaon spectra are described in section~IV,
followed by discussions on the centrality dependence of the kaon
production rates in section~V.
Finally, conclusions are drawn in section~VI.

%%%%%%%%%%%%%%%%%%%%%%%%%%%%%%%%%%%%%%%%%%%%%%%%%%%%%%%%%%%%%%%%%%%%%%%%%%%%
%
%	data analysis
%
\section{Experimental Apparatus and Data Collection}

%	how data were taken?

E--859 was an upgraded experiment from E--802 at the BNL AGS.
The upgrade consisted of an addition of two multi-wire proportional
chambers for triggering purpose to the E--802 spectrometer~\cite{Abb90:Single}
between the Henry Higgins dipole magnet and the time-of-flight
wall, and an implementation of a level-2 trigger~\cite{Zaj91:E859}.
The level-2 trigger linked hits from one of the trigger chambers
and the time-of-flight wall by straight lines to be verified on
the other trigger chamber. Verified combinations were assumed to
originate at the target and particle momenta were obtained.
Particle masses were then obtained from the momenta and
the time-of-flight, and were used for event selection.
Event selection was achieved on-line
by looking up a series of tables pre-loaded in LeCroy CAMAC modules.
For most of the data presented here, the level-2 trigger rejected
events not containing a $K^+$, $K^-$, or $\overline{p}$.
$K^+$ and $K^-$ (or $\overline{p}$) were defined as particles for which
the calculated masses fall in the range of 0.3 $<m<$ 0.7 GeV/$c^2$ and
0.3 $<m<$ 1.5 GeV/$c^2$ , respectively.
Events were also taken with an override on the level-2 trigger
decision to check biases in the level-2 trigger selection criteria. 

E--866 was an upgraded experiment from E--859 designed for measurements
of the Au+Au collisions. The main upgrade was the addition of a
spectrometer for forward angle
measurements~\cite{Ahl98:proton,qm95:Fle,Aki96:QM96}.
The upgrade relevant to this paper included rebuilding the global
detectors for event characterization,
replacing the drift chamber closest to the target,
and an addition of two highly segmented multi-wire
chambers~\cite{Ahl98:kaon_auau,qm95:Fle,Wan96:columbia}
in front of the Henry Higgins magnet.
The addition of the wire chambers was driven by the expected high
multiplicity of central Au+Au events and the need to eliminate
background tracks.
Both wire chambers had drift-time read-out.
Typical resolution of 200 $\mu m$ was achieved for both chambers.

The E--859 data were taken during the 1991 and 1992 Si-beam runs
at the AGS,
and were an improvement over the original E--802 data mainly
due to increased statistics afforded
by the level-2 trigger and the improved acceptance at low momentum.
The data were obtained using a beam of nominal intensity 
$10^6~{\rm ions/s}$ with Al and Au targets of thickness 
817 and 944~${\rm mg/cm^2}$, corresponding to a 3\% and
an 1\% Si interaction rate, respectively.
The data were obtained using 10$^{\circ}$ steps in
spectrometer angle covering the range 5--58$^{\circ}$
and using fields of $\pm 2, \pm 4$~KG 
in the spectrometer magnet~\cite{Col95,Mor94:MIT}.

The E--866 data were taken during the Fall 1993 Au-beam run at the AGS.
The typical beam intensity was
$10^5~{\rm ions/s}$. For the data presented in this paper,
a 975~${\rm mg/cm^2}$ thick Au target was used,
corresponding to a 1.5\% Au interaction rate.
The data were obtained using 5$^{\circ}$ steps in
spectrometer angle covering the range 21--58$^{\circ}$
and using fields of $\pm 2$~KG in the spectrometer magnet.

Both data sets were obtained using a global level-1 trigger that
selected interactions by vetoing on the presence of a beam-charge
ion in the Bull's-eye trigger~\cite{Abb90:Single,Deb98:BE}
downstream of the target, with the additional requirement in the spectrometer
of at least one hit on the time-of-flight wall and a hit on
either of the two trigger chambers.
A plastic scintillator was used in the Bull's-eye
for the Si+A data taking with a threshold, corresponding to a
 cut on the remaining
charge in the projectile of approximately $Z<12.6$.
For the Au+Au data taking, the Bull's-eye was re-designed and rebuilt
as a $\check{\mbox{C}}$erenkov radiator.
The new Bull's-eye had roughly the same geometrical area as the old one,
and used quartz as its radiator.
A Bull's-eye threshold, corresponding to a cut on the remaining
charge in the projectile of approximately $Z<73$,
was used.
The cross-section for events satisfying this cut was measured to be 
5.35~barn. (The total cross-section of Au+Au reaction is
6.8~barn~\cite{Ahl98:kaon_auau,Ahl94:e802_et}.)
Further filtering of data was performed by the level-2 trigger,
when enabled, to enhance the sample of particular particle species
in the data as discussed above.

%%%%%%%%%%%%%%%%%%%%%%%%%%%%%%%%%%%%%%%%%%%%%%%%%%%%%%%%%%%%%%%%%%%%%%
%
%	offline analysis
%
\section{Data Analysis}

In the offline analysis, cuts were applied to the beam counter and 
Bull's-eye data to remove pile-up and to refine the online
Bull's-eye cut. The refined Bull's-eye cut corresponded to
a remaining charge of $Z<11$ in the projectile for the Si+A data, 
and $Z<69$ in the projectile for the Au+Au data, respectively.
The cross-section for events passing this cut was
1.37~barn for Si+Al, 3.68~barn for Si+Au, and 5.23~barn for Au+Au.
For all events satisfying the pile-up and Bull's-eye cuts,
tracks were reconstructed and identified in the spectrometer. 
For the Si+A data, track segments were reconstructed from hits in the 
drift chambers behind the magnet, and projected
through the magnetic field to the front of the spectrometer.
The projections were verified by hits in the front drift chambers
which were refit to give final track parameters.
For the Au+Au data, the track reconstruction behind the magnet was 
done as for Si+A, and an independent track
reconstruction was performed in front of the magnet~\cite{Wan96:columbia}.
The front track reconstruction used the highly segmented wire
chambers to define narrow regions (on order of a few wire pitches,
or 1 $cm$) for track candidates. 
Front drift chamber hits found in these regions were then used,
together with the wire chamber hits with drift time information included,
to obtain track segments in the front.
The track segments in the front and in the back were then
matched through the magnetic field.
A large fraction of background tracks did not have both
the front and the back segments, therefore the matching
algorithm effectively rejected background tracks.
Good tracks were required to project within 2 $cm$ around 
the nominal target position, further reducing background tracks.

The Au+Au particle identification was performed using data from the
time-of-flight wall, since the spectrometer was operated at larger
angles where particle momenta are typically small.
A particle was identified as a kaon if its measured inverse velocity,
$1/\beta$, was within $\pm 3\sigma$ of the expected value and
over $3\sigma$ away from the expected values of other particle types.
The expected values for various particle types were calculated
from the time of flight, track path length up to the time-of-flight wall,
and the corresponding particle masses.
The typical track path length up to the time-of-flight wall
was 660~$cm$. The $\sigma$ was a combination of time-of-flight
resolution and deviation of track trajectory from straight line due to
multiple scattering.
The time-of-flight resolution was measured to be 130 ps.
Hence, kaons were identified up to a momentum of 1.74 GeV/$c$.
For the Si+A data both the time-of-flight wall and the 
segmented $\check{\mbox{C}}$erenkov 
counters~\cite{Abb90:Single} were used for particle identification.
The time-of-flight resolution was measured to be 120 ps for these data.
Kaons were identified up to a momentum of 3 GeV/c.
For both data sets, the contamination of the $K^+$ and $K^-$
samples was found to be below 1\% and 3\%, respectively, averaged over
the spectrometer acceptance~\cite{Wan96:columbia}.

In order to study the centrality dependence of kaon production, the
data were divided into bins in centrality. For the Si+A data the
centrality selection was made using the E--859 Target Multiplicity
Array (TMA), and for the Au+Au data the selection was made using the
zero-degree calorimeter (ZCAL). The Si+A cuts were made in multiplicity
because of a rate-dependent worsening of
the zero-degree calorimeter resolution
during the high beam intensity E--859 run~\cite{Mor94:MIT}.
The zero-degree calorimeter data was used to determine the correlation
between multiplicity and zero-degree energy, $E$,
calculated for events in each multiplicity bin.
The results are shown in Table~\ref{tab:dndy}.
The centrality cuts applied to the data-sets are also listed in 
Table~\ref{tab:dndy} along with the cross-sections for interactions
satisfying the cuts. 

\begin{figure}
\centerline{\epsfxsize=6in\epsfbox[0 200 570 720]{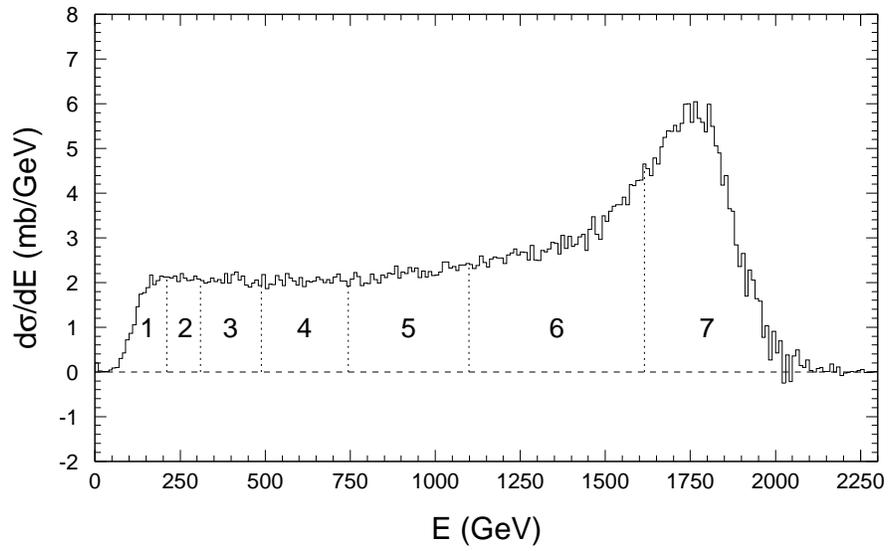}}
\caption{
Calibrated and target-out subtracted zero degree energy spectrum of 
Au+Au interactions from E866.
Centrality bins used in the analysis are labeled 1 through 7.}
\label{fig:zcal_energy}
\end{figure}

Due to the large amount of radiation damage from the high beam intensity
run in E--859, the zero-degree calorimeter was rebuilt for the Au+Au run.
In the Au+Au analysis, the zero-degree energy was calibrated run-by-run
to account for a gradual degradation in the calorimeter response due to
radiation damage~\cite{Wan96:columbia}.
The total degradation over the whole run period was 5\%.
The zero-degree energy resolution was measured to be
1.48 GeV$^{1/2}\cdot \sqrt{E}$.
The calibrated and target-out subtracted zero-degree
energy spectrum for Au+Au events is
shown in Fig.~\ref{fig:zcal_energy}. 
The target-out zero-degree energy was shifted slightly lower
to compensate for the average energy loss of the beam in the target
prior to subtraction.  The target-out interaction rate was 1.1\% for Au+Au and 
1.0\% for Si+A.
Also shown in Fig.~\ref{fig:zcal_energy} as the dotted lines
are the centrality cuts used for
the Au+Au data. The mean zero-degree energy for each bin are
determined from the distribution and results are shown in
Table~\ref{tab:dndy}. 

Total number of identified $K^+$'s and $K^-$'s, after all cuts,
were 120K and 44K for Si+Al, 151K and 69K for Si+Au,
and 28K and 11K for Au+Au.
The statistics were uniformly distributed in available centrality bins.
To obtain semi-inclusive spectra, $K^+$'s and $K^-$'s from events 
satisfying each centrality cut were binned in 
transverse momentum $p_{\perp}$, 
and rapidity $y=\frac{1}{2}\ln\frac{1+\beta_z}{1-\beta_z}$,
where $\beta_z$ is the longitudinal velocity of the particle.
Invariant differential yields, azimuthally averaged, per collision
within a given centrality bin, were obtained according to the formula, 
\begin{eqnarray}
E\frac{d^3 N}{dp^3} 
&=& \frac{d^2 N}{2\pi m_{\perp} dm_{\perp} dy}
 = \frac{d^2 N}{2\pi p_{\perp} dp_{\perp} dy} 	\nonumber \\
&=& \frac{1}{N_{\rm evt}}\cdot
  \frac{\Delta N (y,p_{\perp})}{p_{\perp} 
\Delta p_{\perp} \Delta y}\cdot\frac{1}{A(y,p_{\perp})} ,
\end{eqnarray}
where $N_{\rm evt}$ is the number of events that fell within the given
centrality bin; the transverse mass, $m_{\perp}$, is defined
as $m_{\perp}=\sqrt{p_{\perp}^2+m_0^2}$; $\Delta p_{\perp}$, $\Delta y$
are bin sizes in $p_{\perp}$, $y$, respectively; $\Delta N(y,p_{\perp})$
is the number of kaons in the bin; and $A(y,p_{\perp})$ is the correction
factor. For the Au+Au data, $\Delta p_{\perp} = 50$ MeV/$c^2$
for both $K^+$ and $K^-$; $\Delta y=0.2$ for $K^-$; $\Delta y=0.1$
for $K^+$ in all bins except the most peripheral where $\Delta y=0.2$.
The binning is comparable to that used in the Si+A data set. 
Only particles with momentum 0.3 $< p <$ 1.74 GeV/$c$ were
used in the spectra. 
The correction factor, $A(y,p_{\perp})$, accounted for the
azimuthal acceptance of the spectrometer, particle decays,
multiple-scattering and hadronic interaction losses,
track reconstruction and particle identification
inefficiency~\cite{Wan96:columbia}.
The spectrometer acceptance was calculated from the geometry of the
detectors. Other effects were determined by Monte Carlo
simulations.
The particle decay correction for kaons was found to be roughly the same as
the theoretical momentum-dependent correction because very few kaons
that decayed before the time-of-flight wall were actually reconstructed.
Track reconstruction efficiency depended on hit multiplicity in the
chambers, and was 98\% with the spectrometer at 44$^{\circ}$,
dropping to 95\% at 21$^{\circ}$ for the Au+Au data~\cite{effic_mult}. 
For the Si+Au data, it 
was above 90\% with the spectrometer at 5$^{\circ}$ and was better at larger 
spectrometer angle settings.  For the Si+Al data, it was better than that 
for the Si+Au data at the same spectrometer angle settings.
Track loss due to hadronic interaction for $K^+$ was 1.1\%, independent
of momentum; for $K^-$ it was 2.2\% at high momentum
($>0.6$~GeV/$c$) and increased to 10\% at momentum 0.3 GeV/$c$.
Track losses due to multiple scattering and particle identification
inefficiency
were combined. The net effect was the same for $K^+$ and $K^-$,
and was 4.5\% for high momentum kaons ($>1.4$~GeV/$c$), and increased
to 8.3\% at momentum 0.6~GeV/$c$, and 38\% at momentum 0.3~GeV/$c$.
The level-2 trigger inefficiency was measured to be less than 1\%,
and no correction was applied to the spectra.
Chamber efficiencies were above 95\% for all sense wire planes
in the chambers.
Since not all sense wire planes were required for reconstructed tracks,
the effect of hardware inefficiencies was negligible~\cite{Wan96:columbia}.  
The overall systematic error is estimated at 5\%-10\%.

For the Au+Au data, the spectra are corrected for the energy loss of
particles in the target and in air.
The energy loss in the target was calculated on average
using half thickness of the target and the particle emitting angle.
The energy loss of minimum ionizing particles was about 2 MeV.
The energy loss of a 0.3 GeV/$c$ momentum kaon was 7.5 MeV, corresponding to
a momentum shift of 5\%.

%%%%%%%%%%%%%%%%%%%%%%%%%%%%%%%%%%%%%%%%%%%%%%%%%%%%%%%%%%%
%
%	results and error discussion
%
\section{Results}

\begin{figure}
\centerline{
\epsfxsize=3in\epsfbox[36 0 520 800]{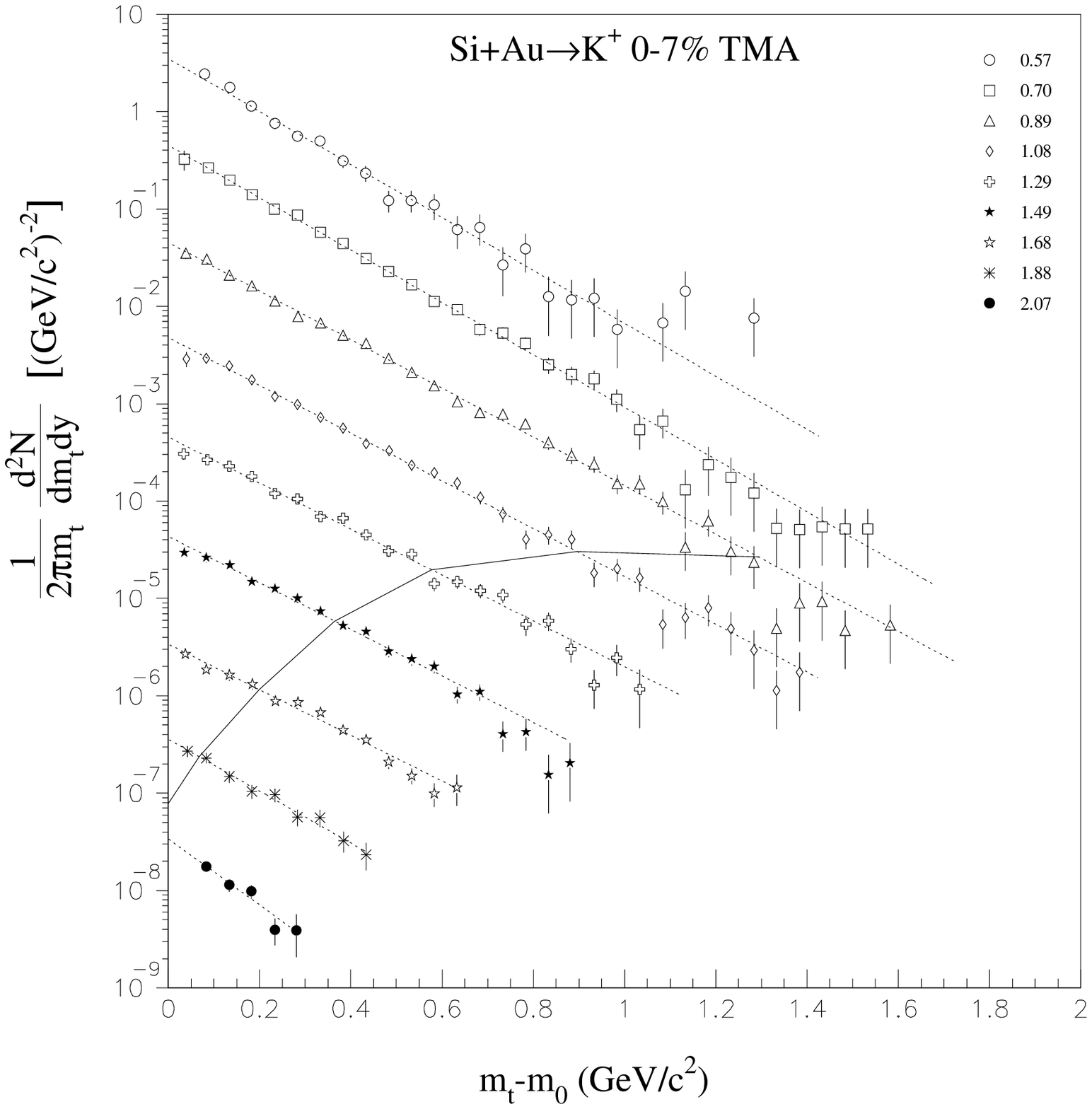}
\epsfxsize=3in\epsfbox[36 0 520 800]{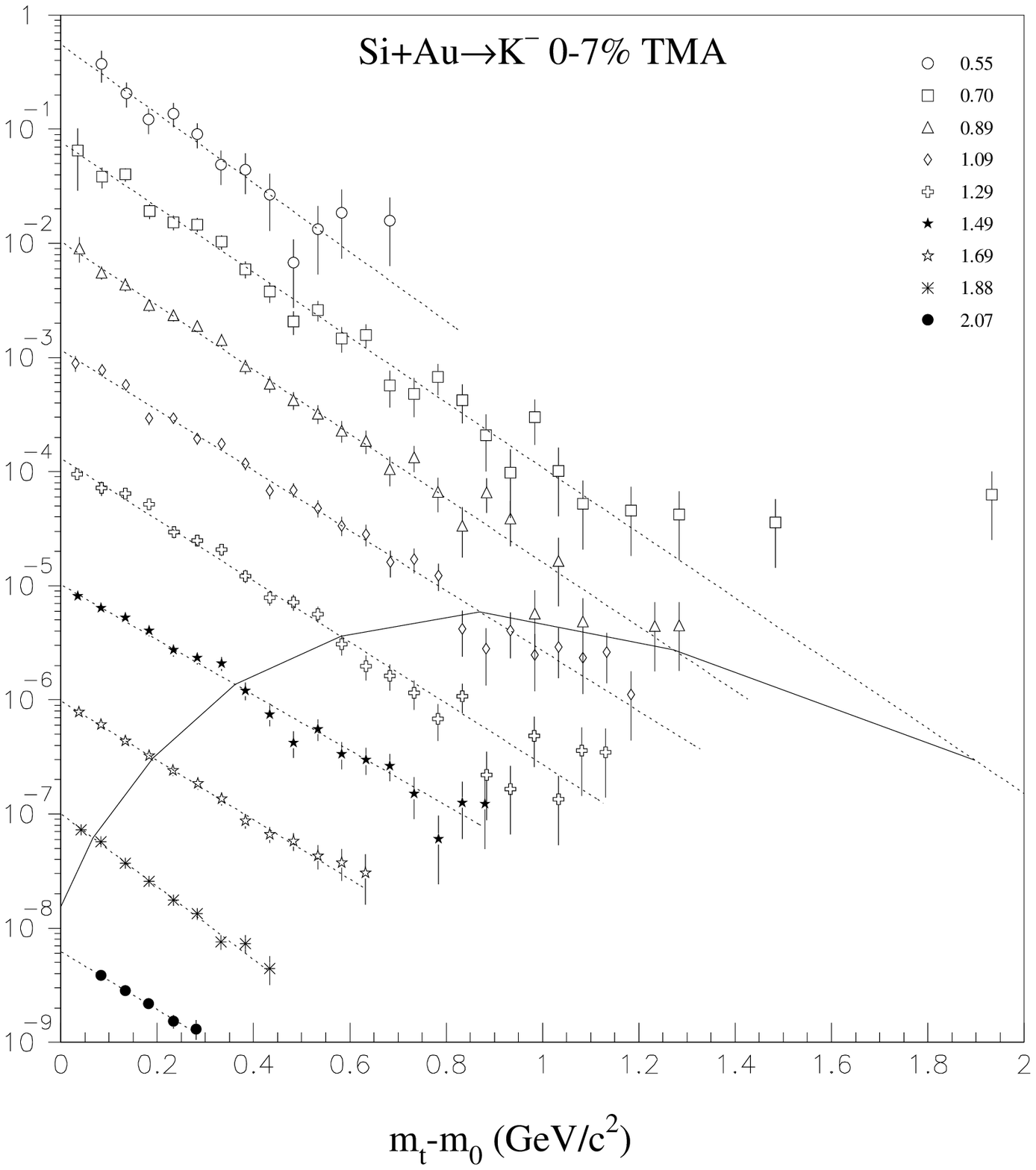}
}
\vspace{1in}
\caption{$K^+$ (left) and $K^-$ (right) transverse mass distributions
in Si+Au for the most central multiplicity bin.
The topmost is absolutely normalized; others are successively divided
by 10 for clarity. 
Data points under the curved line are obtained from particles identified
by both the time-of-flight and $\protect\check{\mbox{C}}$erenkov counters.
Errors shown are statistical only.
Systematic errors were estimated to be 10\%.}
\label{fig:siau_spectra}
\end{figure}

\begin{figure}
\centerline{\epsfxsize=6in\epsfbox[0 120 560 800]{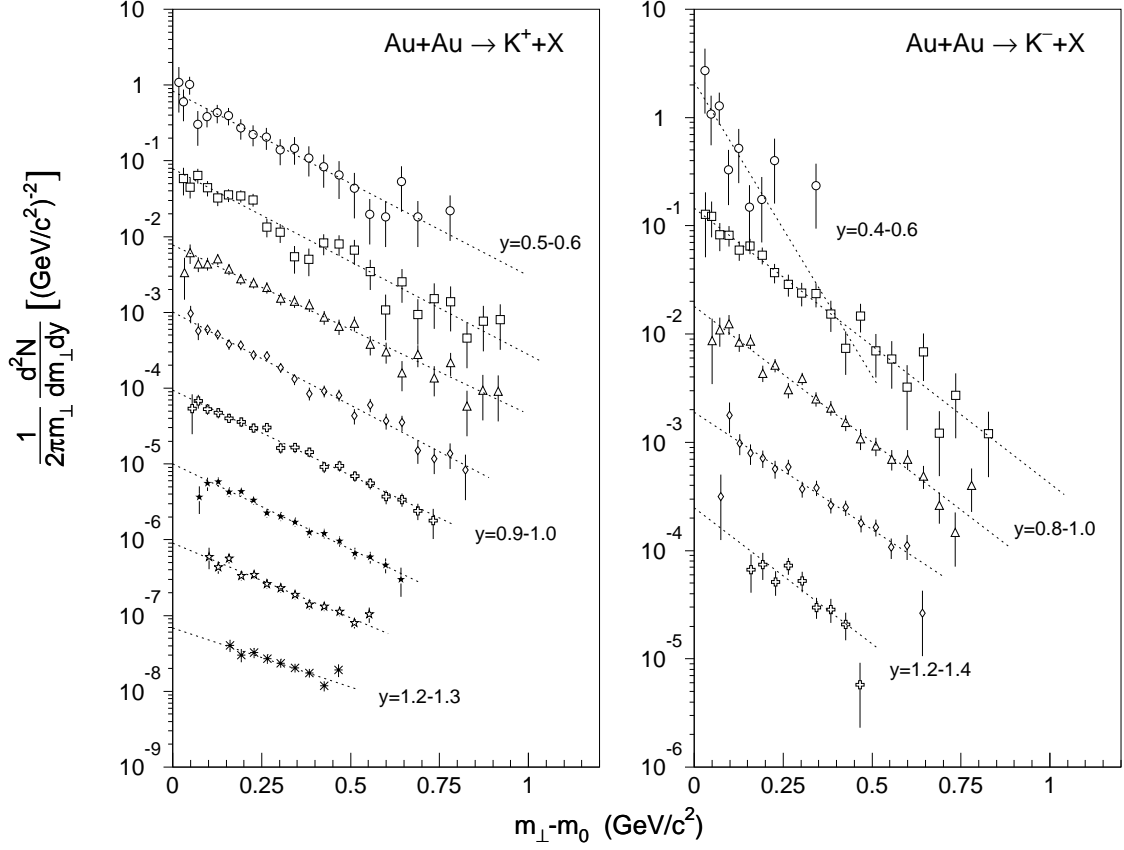}}
\caption{$K^+$ (left) and $K^-$ (right) transverse mass distributions
in Au+Au for the most central zero-degree energy bin.
The topmost is absolutely normalized; others are successively
divided by 10 for clarity. Errors shown are statistical only.
Systematic errors were estimated to be 10\%.}
\label{fig:spectra}
\end{figure}

\begin{figure}
\centerline{
\epsfxsize=3in\epsfbox[20   0 540 660]{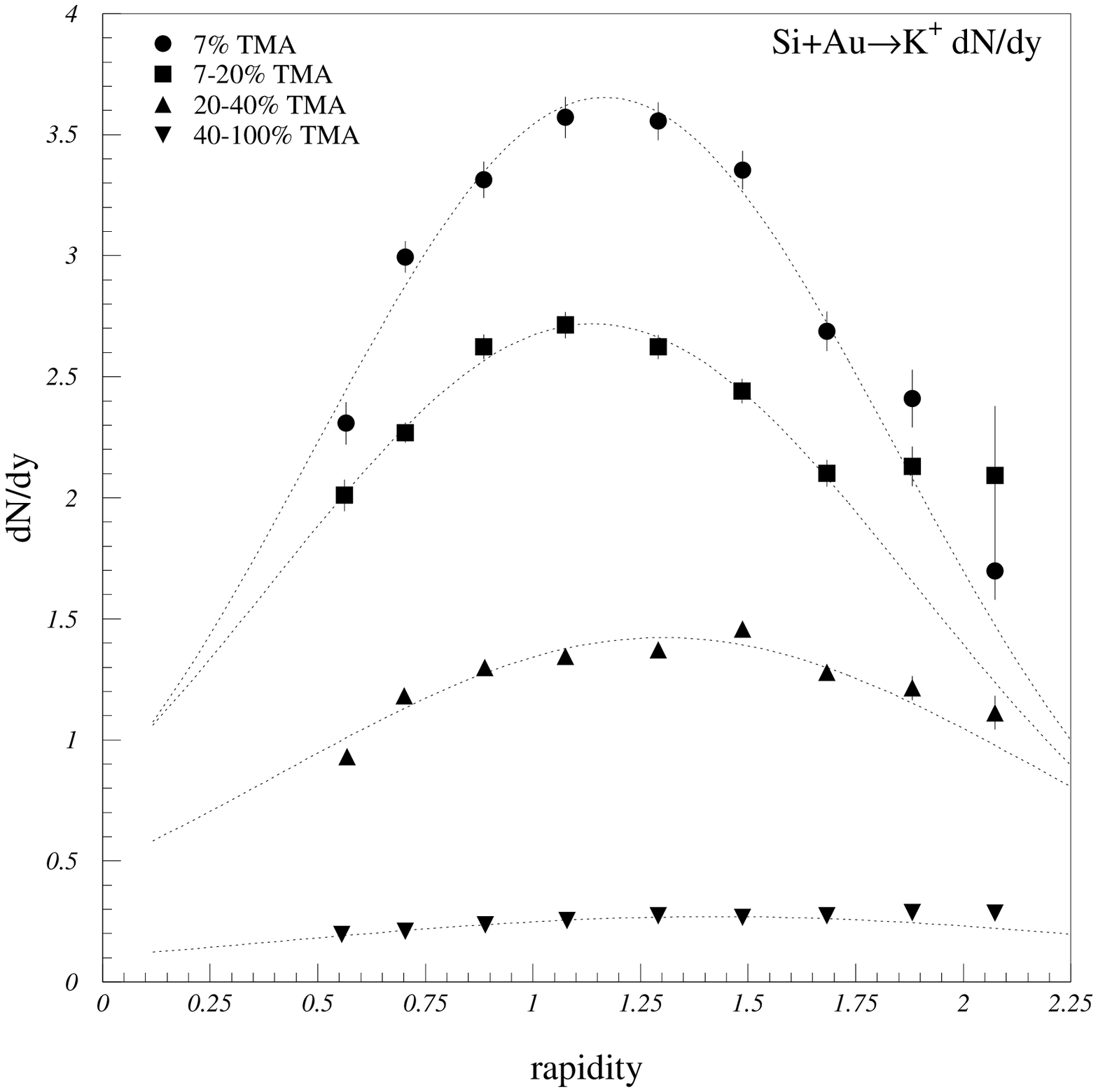}
\epsfxsize=3in\epsfbox[20 142 540 802]{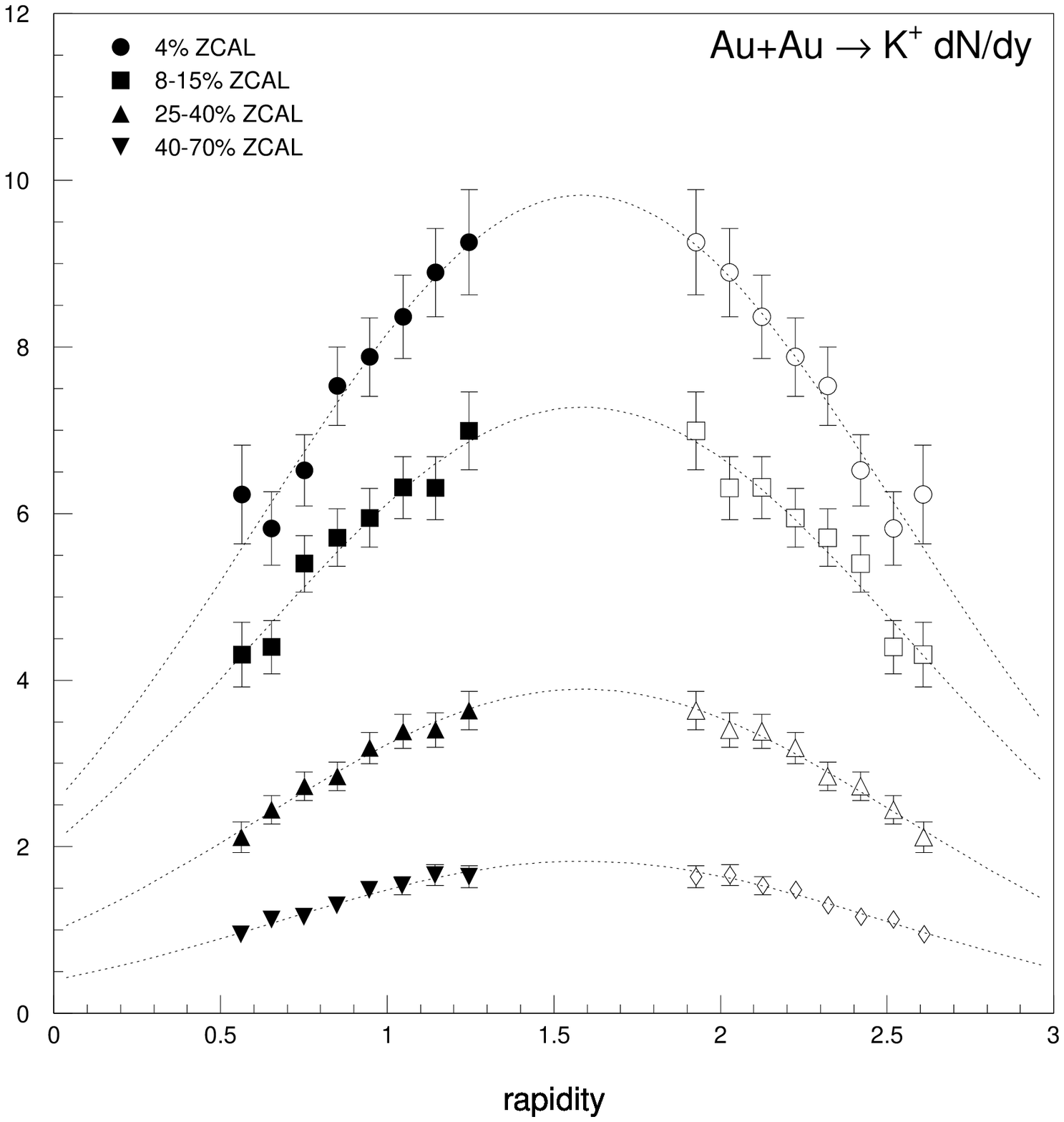}}
\centerline{
\epsfxsize=3in\epsfbox[20   0 540 560]{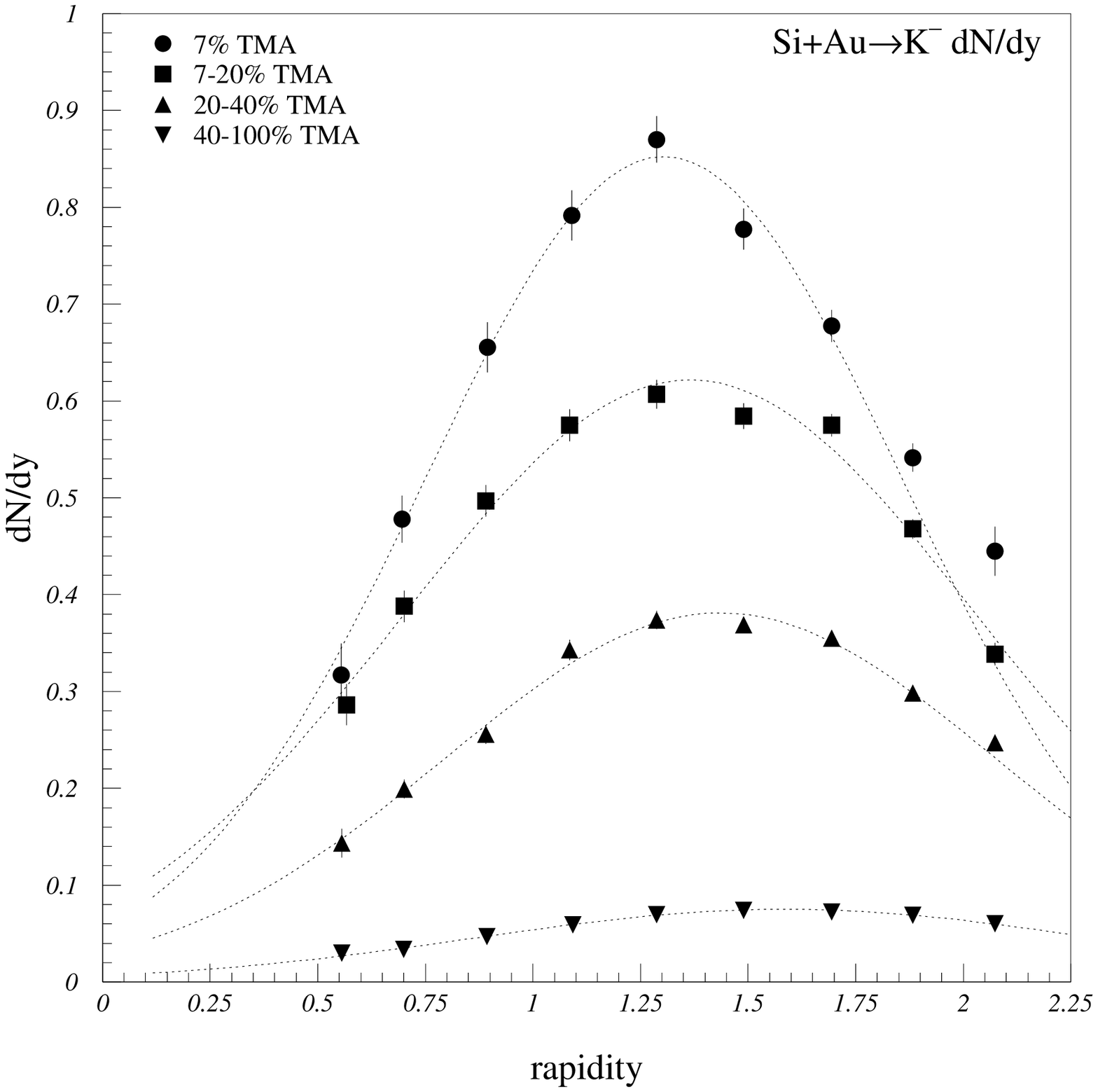}
\epsfxsize=3in\epsfbox[20 142 540 560]{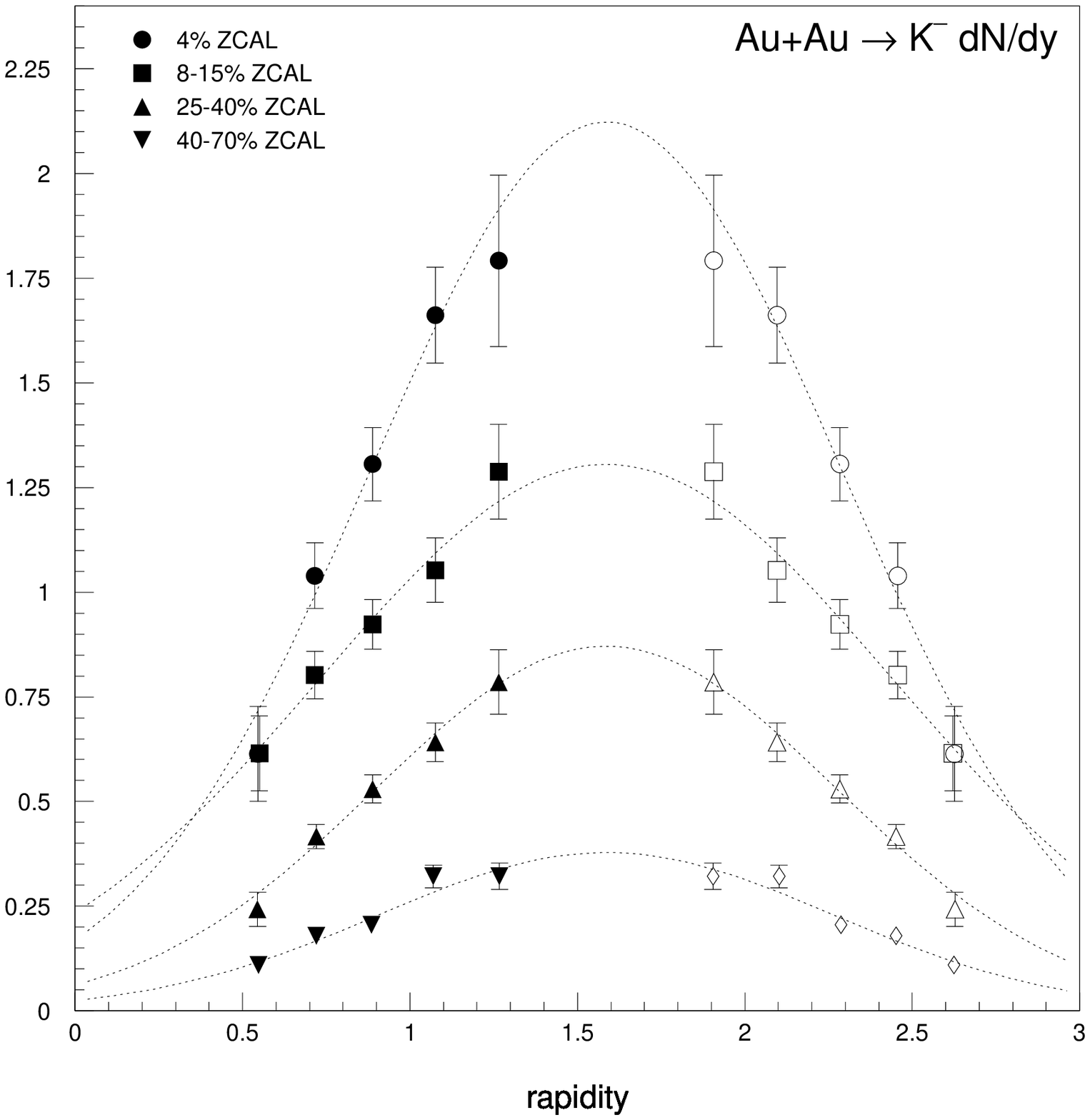}}
\vspace{1in}
\caption{Rapidity distributions of $K^+$ (top) and $K^-$ (bottom)
for Si+Au (left) and Au+Au (right) collisions. 
Filled-in points for Au+Au are data measured; open
ones are data reflected about $y_{NN}$. Errors shown are quadratic
addition of statistical errors and 5\% systematic errors for Au+Au,
and only statistical errors for Si+Au.
There are 10\% systematic errors in the absolute normalization
for both Si+A and Au+Au data.}
\label{fig:dndy}
\end{figure}

Sample sets of kaon transverse mass spectra from Si+Au and Au+Au
collisions, each for the most central bin, are shown in 
Fig.~\ref{fig:siau_spectra} and Fig.~\ref{fig:spectra}, respectively.
All semi-inclusive invariant spectra including those
shown in the figures are well described by an exponential
fall-off with $m_{\perp}$. These spectra are fit to the form
\begin{equation}
E\frac{d^3 N}{dp^3} = \frac{dN}{dy} \cdot 
	\frac{1}{2\pi (m_0T + T^2)} \cdot e^{-(m_{\perp}-m_0)/T} , 
\label{eq:fit}
\end{equation}
which implicitly integrates the spectra over the entire $m_{\perp}$
range, and allows the $dN/dy$ values and the corresponding errors to
be directly extracted from the fit. Results of these fits are
superimposed on the spectra in Fig.~\ref{fig:siau_spectra} and
Fig.~\ref{fig:spectra} as dotted lines.
The obtained $dN/dy$ values
for Si+Au collisions for all four bins in multiplicity measured
in the TMA are plotted in Fig.~\ref{fig:dndy}.
For Au+Au collisions, for clarity, only four of the available seven bins
in zero-degree energy are shown.
The error bars include statistical errors from the fits and 5\%
bin-to-bin systematic errors.
The systematic errors were estimated by comparing yields
in the overlap region between spectrometer settings and by studying the
sensitivity of the spectra to the $m_{\perp}$ intervals used in the fits.
In addition, there was 10\% overall systematic error in normalization. 

For the purposes of this analysis,  rapidity integrated yields
were chosen to characterize the kaon production rates.
The Si+A $dN/dy$ spectra, were fit to
Gaussians and integrated to get the total kaon yields. 
The restricted rapidity coverage of the Au+Au data can be
compensated for by exploiting the required symmetry of the Au+Au
system about the nucleon-nucleon center of mass rapidity ($y_{NN}=1.59$).
If we assume that the Au+Au kaon spectra are well-behaved at
mid-rapidity (as they are in [15]),
then the data are sufficient to obtain an estimate of the total kaon
yield for collisions in each centrality bin. 

The Gaussian fits which account for the
unmeasured tails of the rapidity distributions for both data sets
are shown in Fig.~\ref{fig:dndy} as dotted curves, and were integrated
over all rapidity to estimate the total kaon yields, the results of
which are summarized in Table~\ref{tab:dndy}.
The quoted errors on the kaon yields and $dN/dy$ Gaussian $\sigma$'s
are statistical only. Systematic errors on the Si+A yields were
estimated to be 10\%.
Comparisons with data from the E--866 forward
spectrometer\cite{Ahl98:kaon_auau,Aki96:QM96}
indicate that the assumption regarding the shape of the Au+Au kaon
spectra at mid-rapidity is reasonable, leading to an
estimate that the extrapolation to mid-rapidity introduces a few
percent uncertainty into the present Au+Au total yields. The fractions
of the estimated total $K^+$ and $K^-$ yields in the extrapolated
tails are typically 30\% and 20\%, respectively.
The systematic error on the extracted total kaon yields in Au+Au is
estimated to be of order 10\% by comparing the
extracted yields to alternative fits that have
different shapes for the tails of the distributions.
Using the sum of the measured $dN/dy$
instead of the estimated total yields for Au+Au collisions
does not alter the dependence of the yield on collision centrality
except for a scale factor.
In order from most peripheral to most central collision,
the sum of the $K^+$ $dN/dy$ measured in rapidity range
0.5--1.3 are 
0.29, 1.08, 2.37, 3.56, 4.54, 5.38 and 6.05;
the sum of the $K^-$ $dN/dy$ measured in rapidity range
0.4--1.4 are 
0.076, 0.23, 0.52, 0.74, 0.94, 1.09 and 1.28.

For Au+Au collisions, 
the integrated yields from the present data have been compared with
those described in~\cite{Ahl98:kaon_auau} taken at
a slightly higher beam energy (11.6~A$\cdot$GeV/c).
The $K^+$ and $K^-$ yields of the two measurements were compared
at the 5 centrality cuts from~\cite{Ahl98:kaon_auau}.
The ratios of the yield at the higher energy to that found in
the present work are $1.12 \pm 0.03$ and $1.19 \pm 0.04$
for $K^+$ and $K^-$, respectively, where the quoted
errors are statistical only.
From a parameterized fit to p+p data~\cite{Alb75:pp},
the expected ratios are 1.06 and 1.12
for $K^+$ and $K^-$, respectively.
The differences between the two data sets, when corrected for the
difference in beam energy, are well within the estimated overall
systematic error of 10\% in the measurements.
The results from the 11.1$A$ GeV/$c$ beam energy were used in
the present study because this set contained higher quality data
for kaon production in peripheral collisions.

%%%%%%%%%%%%%%%%%%%%%%%%%%%%%%%%%%%%%%%%%%%%%%%%%%%%%%%%%%%%%%%%
%
%	discussions
%
\section{Discussion}

%	introduce number of participants

In order to discuss the centrality dependence of the kaon yields,
it would be desirable to plot these yields against 
a more intuitive variable
than the zero-degree energy or multiplicity.
In the Wounded Nucleon Model, 
the particle yields in heavy-ion collisions are expected
to increase roughly in proportion to the number of projectile
and/or total participants 
(participating nucleons)~\cite{Abb87:Measurement,Abb92:Global}. Results from
previous Si+A measurements \cite{Abb92:centrality} suggest that kaon
yields within a limited fixed rapidity window increase faster than
linearly with the number of projectile participants 
for Si+Cu and Si+Au, but not for Si+Al collisions.
However, this procedure treats the target and projectile nucleons very
differently for asymmetric collision systems. Consequently, for this work 
we have chosen to use $N_{\rm p}$, the {\em total} number of
participants to characterize collision centrality.

The energy measured in the zero-degree calorimeter results predominantly
from projectile spectators, so the average number of projectile
participants for a given average zero-degree energy can be estimated using the formula  
\begin{equation}
N_{\rm pp} = A_{\rm p} \times \left(1-\frac{E}{E_{beam}}\right) ,
\label{eq:npp}
\end{equation}
where $E$ is taken as the average zero-degree energy in the centrality bin, 
 $A_{\rm p}$ is the number of nucleons in the projectile
(28 for Si and 197 for Au),
and ${E_{beam}}$ is the total kinetic energy of the beam,
383.4 GeV for Si beam and 2000 GeV for Au beam~\cite{energy_loss},
respectively.
For the symmetric Au+Au and nearly symmetric Si+Al collisions,
the total number of
participants is, on average, simply twice the number of projectile
participants. For Si+Au collisions, the total number
of participants has been estimated
by using nuclear geometry with Wood-Saxon density distribution
and the Glauber model to
relate the average number of projectile participants, $N_{\rm pp}$,
to the average number of total participants, $N_{\rm p}$,
in each centrality bin. Results of this calculation
are consistent with those obtained in previous
analyses~\cite{Abb94:Charged,Abb91:zcal_et}.
The estimated total number of participants for the centrality bins
of the three systems are listed in
Table~\ref{tab:dndy}.
The statistical errors on $N_{\rm p}$ are negligible due to the
large number of minimum bias interaction triggers that
were accumulated for each of the three systems.
The systematic errors on $N_{\rm p}$ for the Si+A data
were estimated to be 10\% by using two independent
analyses~\cite{Abb92:centrality,Mor94:MIT}.
For the Au+Au data the systematic errors on $N_{\rm p}$ were estimated to be
8\%~\cite{Wan96:columbia}.  The target-out interaction rate was subtracted from 
the ZCAL distribution 
before $N_{\rm pp}$ was calculated adding negligibly to the estimate of the 
overall systematic error.  The correlation between the ZCAL energy and another 
measure of centrality, the transverse energy, has been analyzed by the E802 
Collaboration and is discussed in 
Refs.~\cite{Abb91:zcal_et,ahle99:E866doublecut,mos93:E866HIPAGS}. 

%%%%%%%%%%%%%%%%%%%%%%%%%%%%%%%%%%%%%%%%%%%%%%%%%%%%%%%%%%%%%%%%%
%
%	physics discussion
%
%	discuss different shapes of Nk vs Np
%

\begin{figure}
\centerline{\epsfxsize=6in\epsfbox[0 100 570 800]{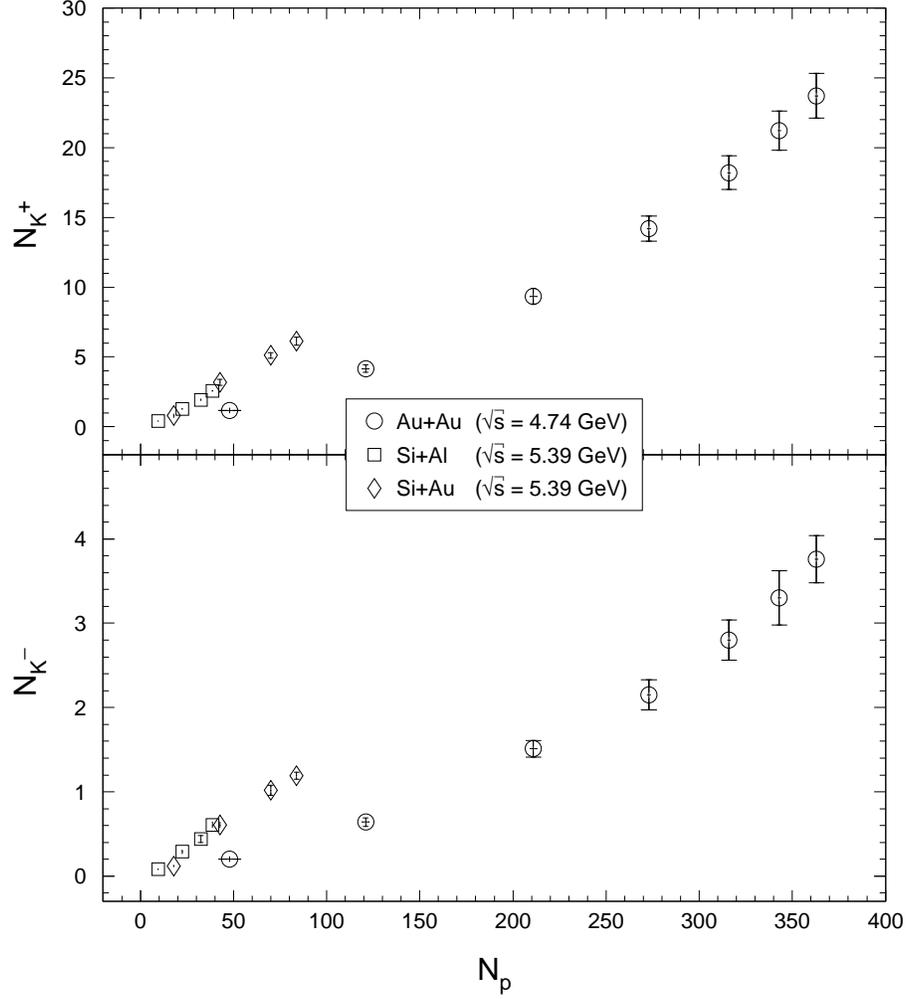}}
\caption{Total $K^+$ (top) and $K^-$ (bottom) yields as
function of the total number of participants, $N_{\rm p}$. 
Errors shown are statistical only. 
Systematic error on $N_{\rm p}$ is 10\% for Si+A data
and 8\% for Au+Au data.
Systematic errors on the yields are 10\% for both Si+A and Au+Au data.
Note that, for clarity, the ordinate scales start from negative values.}
\label{fig:yield}
\end{figure}

\begin{figure}
\centerline{\epsfxsize=6in\epsfbox[0 160 570 800]{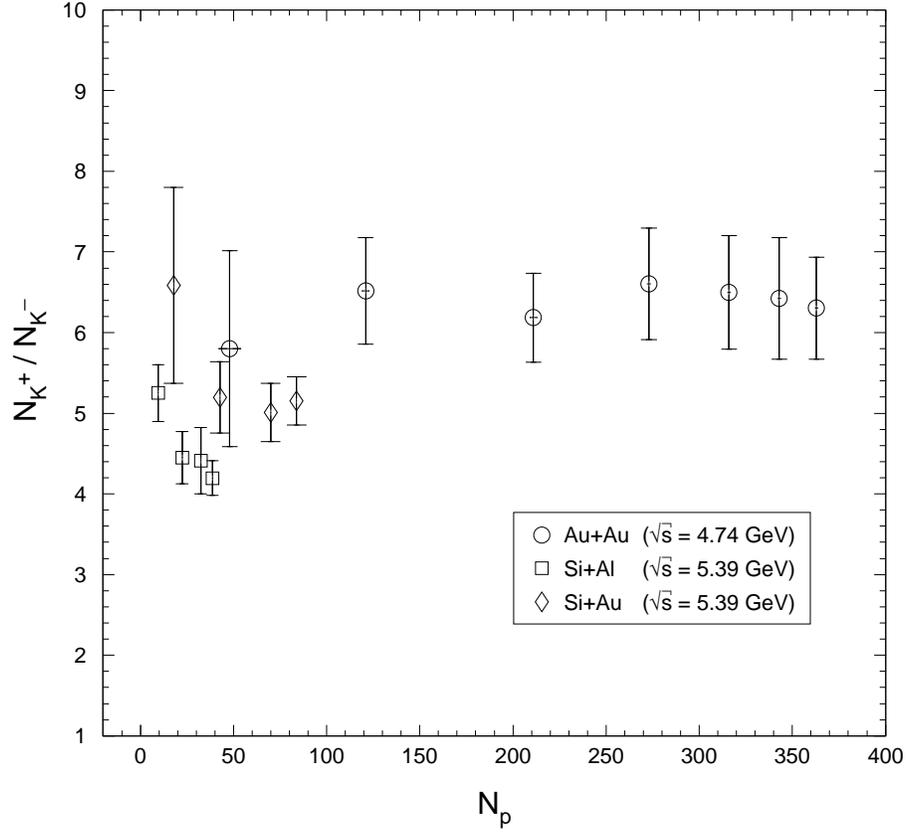}}
\caption{Ratio of $K^+$ to $K^-$ total yield as function of the
total number of participants, $N_{\rm p}$.
Errors shown are statistical only. 
Systematic error on $N_{\rm p}$ is 10\% for Si+A data
and 8\% for Au+Au data.}
\label{fig:ratio}
\end{figure}

The total $K^+$ and $K^-$ yields from Si+Al, Si+Au and Au+Au
collisions are plotted in Fig.~\ref{fig:yield} as a function of $N_{\rm p}$.
One immediate striking feature of this plot is that
the measured kaon yields from Si+A and Au+Au collisions have
very different dependence on $N_{\rm p}$.
The kaon yields from Si+A increase approximately linearly with $N_{\rm p}$
with data from the two targets following approximately the same line.
The kaon yields from Au+Au collisions, in contrast to those from Si+A,
increase faster than linearly with $N_{\rm p}$ as can be clearly seen in
Fig.~\ref{fig:yield}.
On the other hand, in each of the three systems,
the $K^+$ and $K^-$ yields have the same dependence on $N_{\rm p}$,
differing by only a constant scale factor.
This is illustrated in Fig.~\ref{fig:ratio} where the ratios of $K^+$
to $K^-$ yield are plotted against $N_{\rm p}$.
The ratios show a rather weak dependence on $N_{\rm p}$, and are
systematically different for different systems.
(Note that Si+Al and Si+Au are at the same beam energy, and Au+Au
is at a lower beam energy.)
The similarity between $K^+$ and $K^-$ is surprising given that their
production is expected to proceed through very different mechanisms,
resulting in different $dN/dy$ shapes. 
For all three systems discussed, the $K^-$ $dN/dy$ distributions are 
systematically narrower than the $K^+$'s, as seen from Table~\ref{tab:dndy}.

It is of interest to examine the power-law behavior of the kaon yield
versus $N_{\rm p}$: $N_{K} \propto N_{\rm p}^{\alpha}$.  Fits to the Au+Au data yield the 
results $\alpha = 1.54 \pm 0.05$ and $1.52\pm 0.07$, respectively,
for $K^+$ and $K^-$.
The power factors for $K^+$ and $K^-$ are essentially same, consistent with
the constant $K^+$ to $K^-$ ratio, and are considerably larger than 1.

%	discuss different yields in Si+A and Au+Au

\begin{figure}
\centerline{\epsfxsize=6in\epsfbox[0 100 570 800]{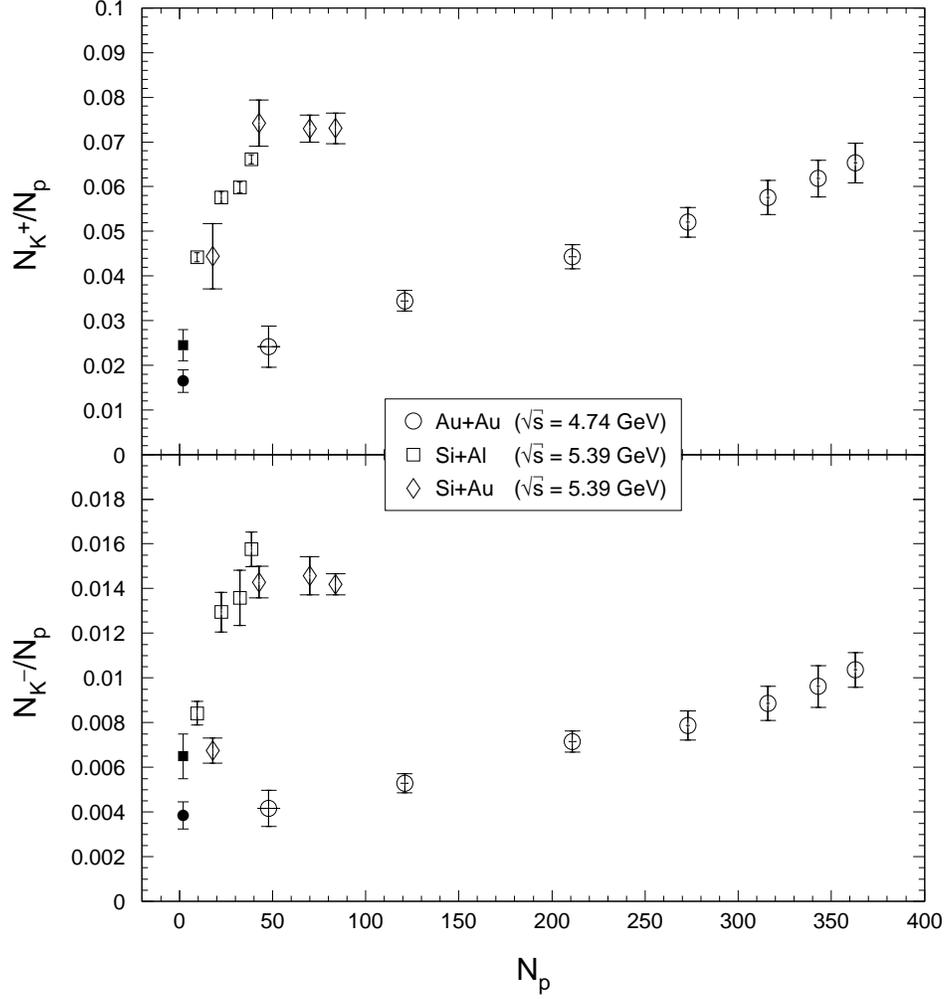}}
\caption{
Total $K^+$ (top) and $K^-$ (bottom) yields per participant
as function of the total number of participants, $N_{\rm p}$. 
Errors shown are statistical only.
Systematic error on $N_{\rm p}$ is 10\% for Si+A data
and 8\% for Au+Au data.
Systematic errors on the yields are 10\% for both Si+A and Au+Au data.
Filled points are those estimated for N+N interactions
at $\protect\sqrt{s}=4.74$ GeV (filled circles) and 5.39 GeV
(filled squares).}
\label{fig:yield_per_part}
\end{figure}

Although the Au+Au kaon yields increase faster than linearly 
with $N_{\rm p}$, the absolute yields fall significantly below
the Si+A yields for the same number of participants ($N_{\rm p}\sim 50$),
which is further shown in Fig.~\ref{fig:yield_per_part} where
the kaon yield per participant is plotted against $N_{\rm p}$.
In order to compare these two data sets,
the difference in beam energy for the two projectiles
must be taken into account.
(The nucleon pair center of mass energy $\sqrt{s}$ is 5.39 GeV for Si+A,
and 4.74 GeV for Au+Au).
The energy dependence of kaon production in heavy ion collisions is unknown;
data from elementary collisions are used to estimate the effect.
$K^+$ and $K^-$ yields in p+p collision at 12~GeV/$c$ were
measured~\cite{Fes79} to be 0.048 and 0.0075, respectively.
Based on a parameterization of p+p data at various
energies~\cite{Alb75:pp}, $K^+$ and $K^-$ yields in p+p should
be scaled, respectively, by 1.27 and 1.53 
from beam momentum 12~GeV/$c$ to 14.6~GeV/$c$,
and by 0.90 and 0.82 from beam momentum 12~GeV/$c$ to 11.1~GeV/$c$.
Therefore, $K^+$ and $K^-$ yields in p+p at 14.6~GeV/$c$ can
be estimated as 0.061 and 0.011, and at 11.1~GeV/$c$ can be estimated as
0.043 and 0.0062. (Typical errors on these yields are 15\%.)
Since there are significant differences in kaon production in p+p, p+n and
n+n interactions~\cite{Gaz91:nn},
the iso-spin averaged kaon yields should be used.
Using the method described in~\cite{Gaz91:nn} and data measured
in p+p at 12 GeV/$c$~\cite{Fes79,Blo74:pp-mult},
the ratios of cross sections for producing $K^+$ in these three elementary
interactions can be estimated as
p+p : p+n : n+n = 1 : 0.81 : 0.62,
and those for producing $K^-$ can be estimated as
p+p : p+n : n+n = 1 : 1.19 : 1.42.
(See appendix of~\cite{Ahl98:kaon_auau} for details.)
Assuming that these ratios do not vary with beam momentum between
11.1 GeV/$c$ and 14.6 GeV/$c$,
and assuming that the heavy-ion collisions are simple superpositions
of nucleon-nucleon (N+N) collisions at the full beam energy,
the $K^+$ and $K^-$ yields per N+N
interaction can be estimated, respectively, as 0.049 and 0.013
for Si+A~\cite{sia_nn} at 14.6 GeV/$c$ and 0.033 and 0.0077 for
Au+Au at 11.1 GeV/c.
The typical errors of these estimates are on the order of 15\%.
These results are shown in Fig.~\ref{fig:yield_per_part} at $N_{\rm p}=2$
as the filled square for the Si beam energy
and the filled circle for the Au beam energy.
Based on these results, the Si+A $K^+$ and $K^-$ yields should be
approximately 40\% and 90\% larger than the Au+Au data if the
heavy-ion collisions were simple superpositions of N+N collisions
at appropriate beam energy. However,
even after compensating for the energy difference the Si+A yields
are still systematically 100\% larger than the Au+Au yields
for the same $N_{\rm p}$.

The above observed differences among the three systems could come from
the different initial nuclear geometry. At the same $N_{\rm p}$, 
the initial geometry of the three systems is by no means the same. 
For the same $N_{\rm p} (\sim 50)$ at which the Au+Au collision
is peripheral, whereas the Si+A collisions are not,
the overlap region in Au+Au collision is less dense than that
in Si+Al or Si+Au collision due to the dilute surfaces of the nuclei.
Therefore, one would naively expect lower kaon yield in Au+Au than
in Si+A collisions at same beam energy, due to fewer successive N+N
collisions per participant in Au+Au than in Si+A collisions.
However, detailed study of the average number of successive N+N
collisions per participant, based on the Glauber model and
the inelastic N+N cross-section 30 mb, does not give
quantitative support to this supposition.

For both Si+Al and Au+Au collisions of all centralities, 
and for peripheral Si+Au collisions,
the kaon yields per participant increase with $N_{\rm p}$,
as seen from Fig.~\ref{fig:yield_per_part}.
In non-peripheral Si+Au collisions, however,
the kaon yield per participant saturates at the value observed
in the most central Si+Al data.
The saturation cannot be explained by the simple nuclear geometry either,
as the average number of successive N+N
collisions per participant, calculated from the Glauber model,
continues to increase with $N_{\rm p}$ in the range of the saturation.

%	Discuss the dependence of Nk vs Np again

Presumably, the kaon yield in {\em very} peripheral heavy ion collisions
should behave just like in N+N.
As shown in Fig.~\ref{fig:yield_per_part}, extrapolations following the
trends of the $K^+$ and $K^-$ yields in both Si+A and Au+Au collisions
are consistent with the N+N values.

%	Discuss strangeness enhancement

\begin{figure}
\centerline{\epsfxsize=6in\epsfbox[0 100 570 800]{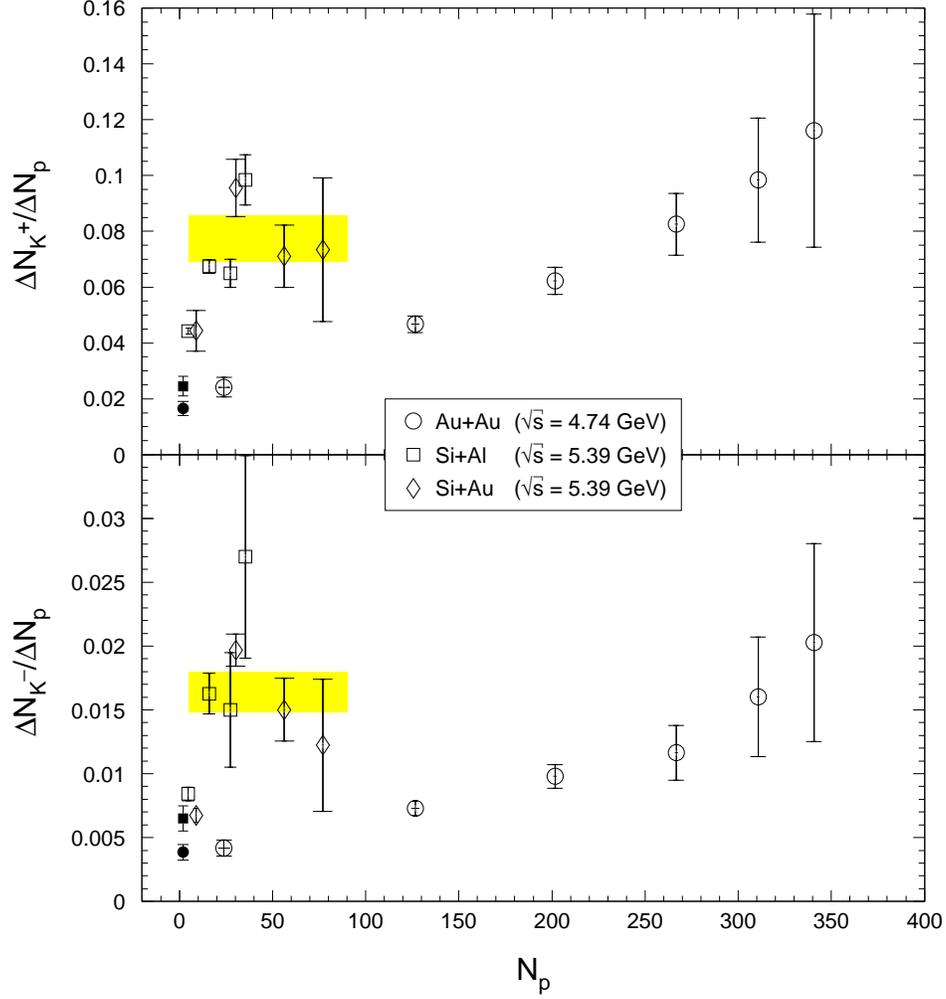}}
\caption{
Differential production rate for $K^+$ (top) and $K^-$ (bottom)
as function of the total number of participants, $N_{\rm p}$. 
Errors shown are statistical only. 
Filled points are the kaon yields per participant in N+N interactions
at $\protect\sqrt{s}=4.74$ GeV (filled circles) and 5.39 GeV
(filled squares).
See text for explanation of the shaded areas and details on the method 
of extracting the differential production rate.}
\label{fig:yield_slope}
\end{figure}

Another way to characterize the dependence of kaon production with
centrality is shown in Fig.~\ref{fig:yield_slope}.
Instead of the average yield per participant displayed in
Fig.~\ref{fig:yield_per_part}, this figure shows the differential rate of
kaon production with increasing $N_{\rm p}$,
{\it i.e.} the slopes extracted from the data in Fig.~\ref{fig:yield}.
Values were extracted separately for each of the three systems.
For the Si data, slopes were found using pairs of adjacent 
points, while for Au+Au data, slopes were obtained from straight-line 
fits to three data points in a row in order to reduce the
statistical uncertainties. The lowest $N_{\rm p}$ point in this figure
was found by combining the lowest data point
in Fig.~\ref{fig:yield} with an assumed point of zero yield at 
zero $N_{\rm p}$.
These results are plotted in Fig.~\ref{fig:yield_slope} at the
corresponding average $N_{\rm p}$ values.
The shaded area shown on each plot
indicates the range of the slope extracted from a straight line fit
to the Si+Al data, the Si+Au data, or the Si data
including both targets shown in Fig.~\ref{fig:yield}.
The kaon yields per participant in N+N interactions are
plotted at $N_{\rm p}=2$ as filled points, the same
as in Fig.~\ref{fig:yield_per_part}.
For the Si-induced interactions, the differential rate of kaon production
rises rapidly and then stays roughly constant at a value considerably
larger than expected from N+N. For Au+Au, the slope increases more
slowly but continuously, and eventually reaches
the value found for central Si+A.
In other words, for the same increase in $N_{\rm p}$,
the increase in kaon yield in central Au+Au is the same as
in central Si+A, in spite of the lower energy of the Au beam
than the Si beam.

Care should be taken in interpreting these results as demonstrating
that each additional participant produces significantly more kaons.
Increasing $N_{\rm p}$ corresponds to decreasing impact parameter.
In a simple geometric model, this implies that all participants have an
increasing probability of multiple collisions.
Also, other effects such as rescattering of baryons and mesons or
the influence of the nuclear medium will be stronger in more central
collisions.
Fig.~\ref{fig:yield_per_part} demonstrates how kaon production
averaged over all participants evolves as a function of centrality
while Fig.~\ref{fig:yield_slope} shows the evolution of the
differential rate of production.

\begin{figure}
\centerline{\epsfxsize=6in\epsfbox[0 100 570 800]{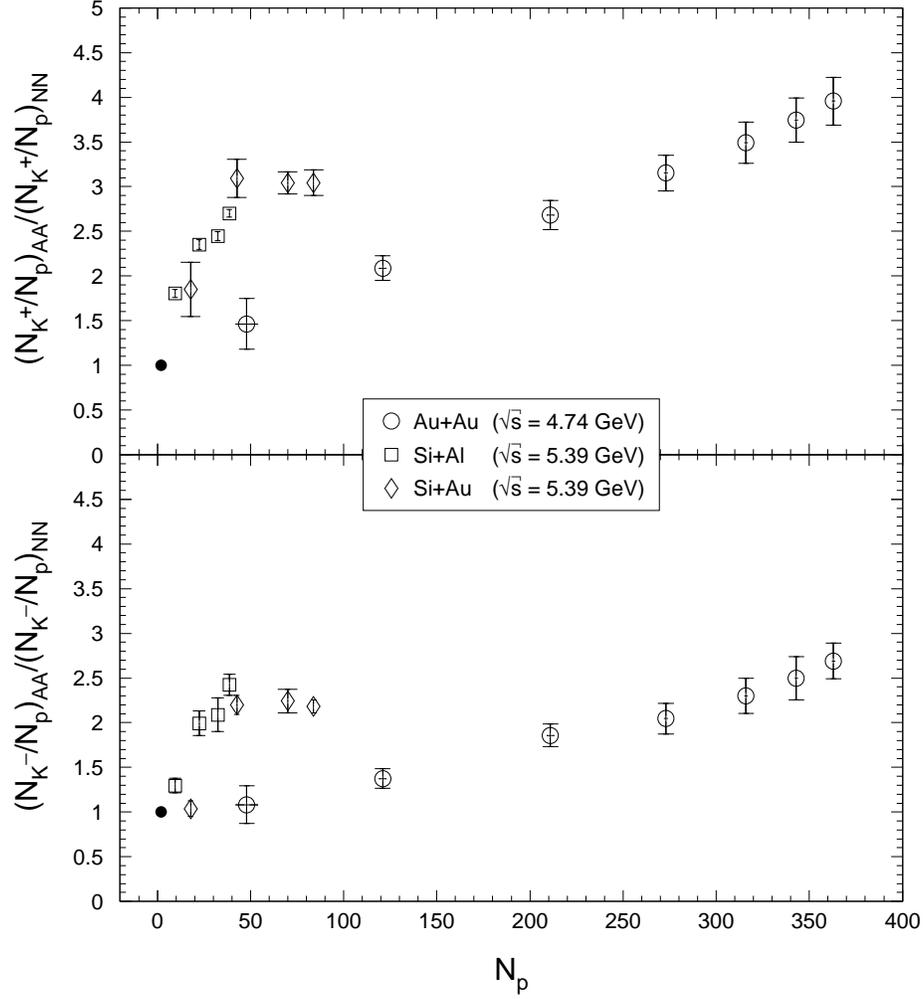}}
\caption{Total $K^+$ (top) and $K^-$ (bottom) yields
per participant, normalized by those in N+N interactions,
as function of the total number of participants, $N_{\rm p}$.
Errors shown are statistical only. 
See text for discussion of the systematic errors.}
\label{fig:yield_over_nn}
\end{figure}

To investigate enhancement of strangeness production
in heavy ion collisions relative to N+N,
the heavy ion data in Fig.~\ref{fig:yield_per_part} are
normalized by the N+N values and re-plotted in Fig.~\ref{fig:yield_over_nn}.
The filled circle at (2,1) in each plot represent the N+N interactions.
The $K^+$ yields in central Si+A collisions are 3 times
those in N+N, a result that is consistent with the previously
observed strangeness enhancement in Si+A collisions\cite{Abb90:Kaon}.
The normalized $K^+$ and $K^-$ yields
per participant in central Au+Au collisions reach higher
values than in central Si+A collisions.
The $K^+$ and $K^-$ yields are enhanced by factors of
 $4.0\pm 0.3 {\rm (stat.)} \pm 0.5 {\rm (syst.)}$ and
 $2.7\pm 0.2 {\rm (stat.)} \pm 0.3 {\rm (syst.)}$, respectively,
in the most central bin of the Au+Au data relative to N+N.
The enhancement factors are only
 $3.0\pm 0.2 {\rm (stat.)} \pm 0.4 {\rm (syst.)}$ and
 $2.2\pm 0.1 {\rm (stat.)} \pm 0.3 {\rm (syst.)}$
in central Si+Au collision for $K^+$ and $K^-$, respectively.
The systematic errors include those in the yields and in $N_{\rm p}$.
Errors in the estimated kaon yields from N+N,
common to all three systems, are not included in the
systematic errors above.
The strangeness enhancement, therefore, is more pronounced
in central Au+Au at 11.1$A$ GeV/$c$ than in central Si+Au
at 14.6$A$ GeV/$c$.
Note that for each of the three systems, there is more enhancement 
in $K^+$ than in $K^-$.

%%%%%%%%%%%%%%%%%%%%%%%%%%%%%%%%%%%%%%%%%%%%%%%%%%%%%%%%%%%%%%%%%%%%%%%%%
%
%	conclusions
%
\section{Conclusions}

In summary, E--859 and E--866 have measured kaon production 
as a function of centrality in Si+Al and Si+Au collisions
at 14.6$A$ GeV/$c$ in rapidity range $0.5<y<2.1$, and in Au+Au collisions
at 11.1$A$ GeV/$c$ in rapidity range $0.5<y<1.3$.
In all three systems with all centralities,
the invariant transverse mass ($m_{\perp}$) spectra can be described by 
a single exponential in $m_{\perp}$ for both $K^+$ and $K^-$;
the rapidity distributions can be characterized by Gaussians,
with the $K^+$ distributions consistently broader than the $K^-$'s.

The kaon yields in Si+A and Au+Au collisions were found to have
different dependencies on the total number of participants.
The kaon yields from Si+A collisions increase approximately linearly with
the total number of participants, whereas the kaon yields from Au+Au
collisions increase faster than linearly.
For the same number of participants ($\sim 50$),
the kaon yields in Au+Au collisions are lower than
in Si+A collisions after the beam energy difference
in kaon production rate in N+N interactions was accounted for.
On the other hand, the $K^+$ and $K^-$ yields in each of the three systems
have quantitatively the same dependence on the total number of participants,
differing only in magnitude by a constant factor,
in spite of the different $dN/dy$ distributions.

Strangeness enhancement is observed as an increase in the slope of the kaon 
yield 
with the total number of participants as well as the yield per participant.
The enhancement in heavy-ion collisions relative to N+N interactions
appears to turn on quickly in Si+A collisions
and saturate at a moderate collision centrality in Si+Au.
The enhancement increases slowly with centrality in Au+Au collisions,
and in central Au+Au collisions, reaches and exceeds the value
found in central Si+A collisions.
The enhancement factor of $K^+$ production are
 $3.0\pm 0.2 {\rm (stat.)} \pm 0.4 {\rm (syst.)}$ and
 $4.0\pm 0.3 {\rm (stat.)} \pm 0.5 {\rm (syst.)}$,
respectively, for the most central 7\% Si+Au collisions and
the most central 4\% Au+Au collisions relative to N+N at the
correponding beam energy.

%%%%%%%%%%%%%%%%%%%%%%%%%%%%%%%%%%%%%%%%%%%%%%%%%%%%%%%%%%%%%%%%%%%%%%%%%%
%
%	Acknowledgement
%
\section{Acknowledgement}

This work was supported by the U.~S.~Department of Energy under contracts
with 
ANL (W-31-109-ENG-38), BNL (DE-AC02-98CH10886), 
Columbia University (DE-FG02-86-ER40281),
LLNL (W-7405-ENG-48), MIT (DE-AC02-76ER03069),
UC Riverside (DE-FG03-86ER40271),
by NASA (NGR-05-003-513), under contract with
the University of California,
by Ministry of Education and KOSEF (951-0202-032-2) in Korea,
and by the Ministry of Education, Science, Sports, and Culture of Japan.

%%%%%%%%%%%%%%%%%%%%%%%%%%%%%%%%%%%%%%%%%%%%%%%%%%%%%%%%%%%%%%%%%%%%%%%%%%
%

%

%%%%%%%%%%%%%%%%%%%%%%%%%%%%%%%%%%%%%%%%%%%%%%%%%%%%%%%%%%%%%%%%%%%%%%%%%%%%%%%%%%
%
%	references
%\begin{thebibliography}{10}

%%%%%%%%%%%%%%%%%%%%%%%%%%%%%%%%%%%%%%%%%%%%%%%%%%%%%%%%%%%%%%%%%%%%%%%%%%
%
%	table
%

\onecolumn

\begin{table}
\caption{Gaussian fit results 
(total yield $N_{K^+}$ and $N_{K^-}$, 
and Gaussian width $\sigma_{K^+}$ and $\sigma_{K^-}$)
of the $K^+$ and $K^-$ $dN/dy$ distributions
for the Si+A data with centrality bins in calibrated target multiplicity
and for the Au+Au data with centrality bins in calibrated zero-degree energy.
The corresponding cross-sections $\sigma$ in mb,
average zero-degree energies $E$ in GeV,
and average numbers of participants $N_{\rm p}$ are also listed.
Quoted errors are statistical only.}
\label{tab:dndy}
\begin{tabular}{c|c|c|c|c|c|c|c|c}
\hline
System		&
Bin		&
$\sigma$	&
$E$		&
$N_{\rm p}$	&
$N_{K^+}$	&
$N_{K^-}$	&
$\sigma_{K^+}$	&
$\sigma_{K^-}$	\\
\hline
		&	% Au+Au data
0--210		&
209		&
155.4		&
363		&
23.7$\pm$1.6	&
3.76$\pm$0.28	&
0.96$\pm$0.06	&
0.71$\pm$0.04	\\
		&
210--309 	&
209		&
259.5		&
343	 	&
21.2$\pm$1.4	&
3.30$\pm$0.32	&
0.97$\pm$0.06	&
0.87$\pm$0.07	\\
		&
309--488 	&
366		&
398.3		&
316	 	&
18.2$\pm$1.2	&
2.80$\pm$0.24	&
1.00$\pm$0.06	&
0.86$\pm$0.06	\\
Au+Au		&
488--744 	&
523		&
616.4		&
273	 	&
14.2$\pm$0.9 	&
2.15$\pm$0.18	&
0.99$\pm$0.05	&
0.80$\pm$0.05	\\
		&
744--1099	&
785		&
926.4		&
211	 	&
9.34$\pm$0.56	&
1.51$\pm$0.10	&
0.96$\pm$0.05	&
0.69$\pm$0.03	\\
		&
1099--1615	&
1569		&
1384.		&
121		&
4.17$\pm$0.27	&
0.64$\pm$0.05	&
0.91$\pm$0.05	&
0.68$\pm$0.03	\\
		&
1615--2300	&
1569		&
1754.		&
48		&
1.16$\pm$0.17	&
0.20$\pm$0.03	&
0.98$\pm$0.12	&
0.67$\pm$0.07	\\
\hline
		&	%most central Si+Au data
$>$115		&
259		&
21.8		&
83.9		&
6.13$\pm$0.29	&
1.19$\pm$0.04	&
0.67$\pm$0.03	&
0.56$\pm$0.02	\\
Si+Au		&
85--115		&
481		&
51.7  		&
70.0		&
5.11$\pm$0.21	&
1.02$\pm$0.06	&
0.75$\pm$0.03	&
0.67$\pm$0.04	\\
		&
50--85		&
741		&
149.6		&
42.7		&
3.17$\pm$0.22	&
0.61$\pm$0.03	&
0.89$\pm$0.06	&
0.64$\pm$0.03	\\
		&	% most periph Si+Au data
0--50		&
2222		&
278.8		&
17.8		&
0.79$\pm$0.13	&
0.12$\pm$0.01	&
1.16$\pm$0.19	&
0.69$\pm$0.05	\\
\hline
		&	% most central Si+Al data
$>$48.5		&
97		&
118.3		&
38.7		&
2.56$\pm$0.04	&
0.61$\pm$0.03	&
0.82$\pm$0.01	&
0.71$\pm$0.04	\\
Si+Al		&
35.5--48.5	&
180		&
161.8		&
32.4		&
1.94$\pm$0.04	&
0.44$\pm$0.04	&
0.90$\pm$0.02	&
0.71$\pm$0.04	\\
		&
22.5--35.5	&
277		&
229.8		&
22.4		&
1.29$\pm$0.03	&
0.29$\pm$0.02	&
0.90$\pm$0.02	&
0.73$\pm$0.04	\\
		&	% most periph Si+Al data
0--22.5		&
831		&
318.2		&
9.5		&
0.42$\pm$0.01	&
0.080$\pm$0.005	&
1.04$\pm$0.03	&
0.78$\pm$0.05	\\
\hline
\end{tabular}
\end{table}

%%%%%%%%%%%%%%%%%%%%%%%%%%%%%%%%%%%%%%%%%%%%%%%%%%%%%%%%%%%%%%%%%%%%%%%%%%
%
%	figure captions
%

\end{document}